# Surface Conduction and Electroosmotic Flow around Charged Dielectric Pillar Arrays in Microchannels

Keon Huh[1], So-Yoon Yang[2], Jae Suk Park[1,3], Jung A Lee[1], Hyomin Lee[4†] and Sung Jae Kim[1,3,5†]

[1]*Department of Electrical and Computer Engineering,*
*Seoul National University, Seoul, 08826, Republic of Korea*

[2]*Department of Electrical Engineering and Computer Science,*
*Massachusetts Institute of Technology, Cambridge MA 02139, USA*

[3]*Inter-university Semiconductor Research Center,*
*Seoul National University, Seoul, 08826, South Korea*

[4]*Department of Chemical and Biological Engineering,*
*Jeju National University, Jeju, 63243, Republic of Korea*

[5]*Nano Systems Institute,*
*Seoul National University, Seoul, 08826, South Korea*

[†]Correspondence should be addressed to Prof. Hyomin Lee and Prof. Sung Jae Kim
by e-mail: (HLee) fluid@jejunu.ac.kr; (SJKim) gates@snu.ac.kr

**ABSTRACT**

Dielectric microstructures have been reported to have negative influences on perm-selective ion transportation because ions do not migrate in areas where the structures are located. However, the structure can promote the transportation if the membrane is confined to a microscopical scale. In such scale where the area to volume ratio is significantly large, the primary driving mechanisms of the ion transportation are transited from electro-convective instability (EOI) to surface conduction (SC) and electroosmotic flow (EOF). Here we provide rigorous evidence on how SC and EOF around the dielectric microstructures can accelerate the ion transportation by multi-physics simulations and experimental visualizations. The microstructures further polarize the ion distribution by SC and EOF so that ion carriers can travel to the membrane more efficiently. Furthermore, we verified, for the first time, that the arrangements of microstructures have a critical impact on the ion transportation. While convective flows are isolated in crystal pillar configuration, the flows show elongated pattern and create an additional path for ion current in the aligned pillar configuration. Therefore, the fundamental findings of the electrokinetic effects on the dielectric microstructures suggest an innovative application in an energy-efficient micro/nanofluidic devices.

**Abbreviations**

ICP: Ion concentration polarization

EDL: Electrical double layer

SC: Surface conduction

EOF: Electroosmotic flow

EOI: Electroosmotic instability

OLC: Overlimiting current

IDZ: Ion depletion zone

IEZ: Ion enrichment zone

AIDZ: Additional ion depletion zone

AIEZ: Additional ion enrichment zone

# 1. Introduction

Perm-selective ion transportation through the nanoporous membrane has gained significant attention for studying fundamental electrokinetics and developing innovative engineering applications based on the fundamentals. Traditional electrodialysis[1-4] system is one of the representative platforms where the perm-selective ion transport leads to a distinguishable overlimiting conductance[5, 6] due to concentration polarization phenomenon and an efficient salt removal for desalination application[7-10]. The working principle of electrodialysis is that source water injected into serial stacks of cation-exchange- and anion-exchange-membrane is separated into waste water and fresh water with a perpendicularly applied electric field across the membrane stacks[11, 12]. Thus, spacers should be installed between the membrane stacks to prevent any physical- and electrical- contact. However, it has been reported that the dielectric spacers have a negative impact on the ion transport due to the current shadow effect[13]. Ions do not migrate in areas where the spacers are located and therefore the spacers lead to hinder the ion transportation, although they enhance mass transfer *via* mixing promotions[11] or local flow redistributions [14]. While a conductive spacer[15-17] has also been utilized as a conduction promoter[11, 18, 19], it is difficult to manufacture the conductive spacer because it can easily lose conductivity by physical deterioration.

Instead, here we suggest avoiding the current shadow effect by reducing the characteristic length scale of the system, while the dielectric spacers remain used. It has been reported that the length scale is a critical factor to distinguish the driving mechanisms of perm-selective ion transportation[5, 20, 21]. Depending on the scale, the mechanisms are categorized into (i) surface conduction (SC)[5, 22, 23] with $O(1)$ μm constriction, (ii) electroosmotic flow (EOF)[24-27] with $O(10)$ μm scale and (iii) electro-osmotic instability (EOI)[28-36] above $O(100)$ μm scale. The length scale of typical electrodialysis system lies in the EOI regime[37] so that reducing convection by the spacer would lower the ion transport. However, microfluidic confinement can help enhance the ion transportation because the length scale of such system is typically within the SC or EOF dominant environment[23, 38-40]. As expected, preliminary measurements with or without micro dielectric pillar structures (which mimicked the spacer in the electrodialysis system) can enhance overlimiting conductance as shown in Figure 1(a). Note that overlimiting conductance can be measured by calculating the slope of current in overlimiting regime. In such confinement, the electrical double layer (EDL)[41-44] around the dielectric pillar with negatively charged surface contributes to additional ion transport by SC (direct conduction within EDL) and associated EOF (initiating additional flow originated from EDL) as shown in Figure 1(b).

First of all, SC caused by the EDL around the dielectric pillar results in an additional local ion concentration polarization (ICP) phenomenon around each pillar as shown in Figure 1(b). Note that conventional ICP can be described as the formation of polarized electrolyte concentration at both sides of a nanoporous membrane under an external electric field; low concentration zone (*i.e.* an ion depletion zone (IDZ)) at an anodic side and high concentration zone (*i.e.* an ion enrichment zone (IEZ)) at a cathodic side of cation-selective membrane[45, 46]. When the pillars are installed inside IDZ, additional ion depletion zone (AIDZ), which decreases the electrical conductance, forms at the back of each pillar (*i.e.* the location facing the bulk) as shown in Figure 1(c). In contrast, additional ion enrichment zone (AIEZ) which increases the electrical conductance appears at the front of each pillar (*i.e.* the location facing the cation-selective surface). Since the dielectric pillars are arranged under appropriate separation distance, a number of AIEZ formed inside IDZ can create an additional current path, while the current barely flows through IDZ in the case of "without pillar". Thus, the AIEZ resulting from SC increases the electrochemical efficiency of the system by directly serving the abundant charge-carrier toward the nanoporous membrane. Secondly, EOF is the motion of fluid that appears around each dielectric pillar so that EOF can drag and supply the charge-carriers nearby the dielectric pillars, to AIEZ. Therefore, we are expecting the combined contribution of SC and EOF caused by the dielectric pillars enhances the additional ion current by forming multiple AIEZs.

The dominance of SC and EOF are reported to vary with (i) surface charge of a substrate and (ii) Debye length around the substrate (*i.e.* bulk concentration)[5]. Thick Debye length and high surface charge drive the system to a SC-dominant regime and pillars with thin Debye length and low surface charges let the system be EOF-dominant regime. When the surface charge converges to zero, the electrokinetic effect of the dielectric pillar is vanished, which has been the research subject of the non-conductive spacer in the electrodialysis system [13]. Therefore, confirming the formation of AIEZs at the front of dielectric pillars and electroosmotic flow around the dielectric pillars would be a crucial evidence of SC- and EOF-governed regimes in the case of micro-confined platforms. In this study, we rigorously investigated the evidences by multi-physics simulations and experiment using micro/nanofluidic platforms[24, 47, 48]. Individual roles of SC and EOF are extracted and compared with various electrokinetic parameters such as an EDL thickness[49, 50], a surface charge of dielectric pillar and an external electric field. Furthermore, we present the effects of physical configuration of dielectric pillar arrays on overlimiting conductance, which has not been typically considered in previous studies. Thus, the conclusion derived from this analysis would provide a general strategy for designing highly energy-efficient nanoporous membrane platform.

## 2. Materials and Methods

### 2.1 Numerical Methods

#### 2.1.1. Numerical domain

The microfluidic confinement membrane system with micro dielectric pillars can be represented as a schematic domain as shown in Figure 2. When an electric field is applied across stationary bulk to cation-selective surface, the direction of electro-migrations are different for cation and anion, but anions are not replenished from the opposite side of the membrane. Thus, the total salt concentration in the vicinity to the anodic side of membrane would be largely depleted to form an ion depletion zone (IDZ) and the concentration at the cathodic side would be enriched to form an ion enrichment zone (IEZ). This polarization across the membrane is called ion concentration polarization (ICP)[51, 52]. Note that ICP is different from "concentration polarization" phenomenon in an electrodialysis system because they have their unique major driving mechanisms depending on the critical length scale[20, 22, 53]. Dielectric pillar array (diameter = $2a$ and inter-pillar distance is $b_1$ in $y$-direction and $b_2$ in $x$-direction) is installed apart from the cation-selective surface at $b_3$. The cation-selective surface is assumed to have an ideal perm-selectivity so that only cation-flux can pass through.

We used general form PDE module of COMSOL to directly input our dimensionless governing equations and boundary conditions. Other predefined modules (e.g. electrostatics, transport of diluted species, etc) are inconvenient to implement the non-dimensional problem. Using the general form PDE module, any type of governing equations and boundary conditions can be implemented and solved on COMSOL. Figure 3 showed our computational domain. To reduce computational costs, we employed two axes of symmetry at $y = 0$ and $y = 0.25$ and only considered three columns of pillar. Under a dimensionless length scale, the $x$- and $y$-directional domain length were set to be 1 and 0.25. The pillar radius, $a$ was 0.1. The parameters about pillar to pillar interval were $b_1 = 0.5$, $b_2 = 0.3$ and $b_3 = 0.2$. For a KCl electrolyte, $Pe^0 = 0.22$.

**2.1.2. Governing equations**

In order to simultaneously calculate ICP and surface conduction/electroosmotic slip on dielectric pillars, we employed local electroneutral model established by Rubinstein and Zaltzman[33, 54] and asymptotic model for SC and EOF established by Schnitzer and Yariv[55]. Referring to their work, this section briefly introduced the model.

The dimensionless Poisson equation describes the conservation of ionic charge.

$$-2\lambda_0^2 \nabla^2 \phi = c_+ - c_- \tag{1}$$

where $\lambda_0$, $\phi$, $c_+$ and $c_-$ are the Debye length, electrical potential, cation concentration and anion concentration, respectively. We chose normalizing scales as follows; (i) distance from cation-selective surface to stationary bulk as a length scale, (ii) thermal voltage as an electrical potential scale, (iii) bulk concentration as a concentration scale. As $\lambda_0 \to 0$, the Poisson equation implied that

$$0 = c_+ - c_-. \tag{2}$$

In other words, we can assume an electroneutral domain. Thus, cation and anion concentrations can be denoted as

$$c_+ = c_- = c. \tag{3}$$

Using above electroneutral relation, the dimensionless Nernst-Planck equation to describe the transport phenomena of ionic species were

$$\nabla \cdot \left( -\nabla c - c \nabla \phi + Pe^0 c \mathbf{u} \right) = 0 \tag{4}$$

for a cationic species, and

$$\nabla \cdot \left(-\nabla c + c\nabla\phi + Pe^0 c\mathbf{u}\right) = 0 \tag{5}$$

for an anionic species. The velocity scale ($\tilde{U}_0$) was set to be

$$\tilde{U}_0 = \frac{\tilde{\varepsilon}\tilde{V}_T}{\tilde{\eta}\tilde{L}} \tag{6}$$

where tilde symbol (~) means dimensional variable, $\tilde{\varepsilon}$ is the electrical permittivity of water, $\tilde{V}_T$ is the thermal voltage scale, $\tilde{\eta}$ is the viscosity of water and $\tilde{L}$ is the distance from cation-selective surface to stationary bulk. $Pe^0$ is the material Peclet number which is defined by

$$Pe^0 = \frac{\tilde{\varepsilon}\tilde{V}_T}{\tilde{\eta}\tilde{D}} \tag{7}$$

where $\tilde{D}$ is the diffusivity of dissolved ion. Note that we assumed the same cation and anion diffusivity ($\tilde{D}_+ = \tilde{D}_- = \tilde{D}$). This assumption is valid in KCl electrolyte. Adding equation (5) to equation (4),

$$\nabla \cdot \left(-\nabla c + Pe^0 c\mathbf{u}\right) = 0. \tag{8}$$

This implied mass conservation. Substituting equation(5) from equation (4),

$$\nabla \cdot \left(-c\nabla\phi\right) = 0 \tag{9}$$

This equation implied electrical current conservation. The flow field, **u** and pressure, $p$ was governed by following dimensionless Stokes equations and continuity equation of

$$\mathbf{0} = -\nabla p + \nabla^2 \mathbf{u} \tag{10}$$

$$\text{and } \nabla \cdot \mathbf{u} = 0 \tag{11}$$

Since the electroneutral domain was assumed by $\lambda_0 \rightarrow 0$, the volumetric body force term in the Stokes equations was omitted.

### 2.1.3. Boundary conditions

The boundary conditions on stationary bulk were

$$c = 1, \quad \phi = V \quad \text{and} \quad p = 0 \tag{12}$$

where $V$ is the applied voltage. On cation-selective surface,

$$\ln c + \phi = 0, \quad -\frac{\partial c}{\partial x} + c\frac{\partial \phi}{\partial x} = 0 \quad \text{and} \quad \mathbf{u} = \mathbf{0}. \tag{13}$$

Above conditions, the first one refers that electrochemical potential set to be zero. The second is zero-anion flux through cation-selective surface and third one is no-slip boundary condition. On the dielectric micropillar surfaces, the surface conduction was described by the flux-matching condition between normal and tangential flux referred to Schnitzer and Yariv's analysis[55]. Note that Schnitzer and Yariv analyzed the surface conduction in spherical coordinate system, so that their analysis should be transformed into local cylindrical coordinate system as denoted in Figure 3. For cationic species,

$$-\frac{\partial c}{\partial r} - c\frac{\partial \phi}{\partial r} = Du \frac{\partial^2}{\partial \theta^2}(\phi + \ln c) \tag{14}$$

where $r$ and $\theta$ are normal and tangential components of which origin is at each center of micropillar. $Du$ is the Dukhin number of micropillar defined as

$$Du = \left(1 + 2Pe^0\right)\lambda_0 q_s \tag{15}$$

where $\lambda_0$ is the dimensionless Debye length normalized by pillar radius and $q_s$ is the dimensionless surface charge density of pillar. By the aid of above matching boundary condition (equation (14)), the contribution of surface conduction can be expressed by using $Du$. In addition, this approach is helpful to reduce computational cost significantly because the ion transport within electrical double layer is treated as simple boundary condition rather than coupled differential equation of Poisson and Nernst-Planck equations. Due to highly charged micropillar surface, the surface flux of anion was zero. This zero surface flux led to

$$-\frac{\partial c}{\partial r} + c\frac{\partial \phi}{\partial r} = 0. \tag{16}$$

The electroosmotic slip on micropillar surfaces was described by following slip condition

$$u_\theta = \zeta \frac{\partial \phi}{\partial \theta} - 4\ln\left(\cosh\frac{\zeta}{4}\right)\frac{\partial \ln c}{\partial \theta}. \tag{17}$$

The first and second terms are electroosmosis and diffusioosmosis, respectively. The zeta potential ($z$) can be evaluated by following Grahame equation,

$$q_s = 2\sqrt{c}\sinh\frac{\zeta}{2}. \tag{18}$$

Equations (8) – (18) were solved by commercial finite element software, COMSOL Multiphysics.

### 2.1.4. Baseline comparison ($Pe^0 = 0$)

We performed additional simulation at $Pe^0 = 0$ as shown in Figure 4 below. Since $Pe = 0$ means no convection, AIDZ and AIEZ are clearly observable and tend to vertically spread more by comparing the cases of $Pe^0 = 0$ and $Pe^0 \neq 0$. Note that the concentration distribution distorted more in (b) the aligned configuration because the role of convection is more dominant than (a) the crystal configuration.

### 2.2 Experimental Methods

### 2.2.1 Device fabrication

To investigate the effect of the dielectric pillar, the design of the microchannel is shown as in Figure 5. The pillar array (15 μm radius and inter-pillar distance of 95 μm and 100 μm) were installed inside the main microchannel. We examined two types of pillar configuration; crystal- and aligned- configuration. Historically, pillar structures have been utilized as immuno-bead trappers in ICP biosensor platform[56, 57]. In this platform, the pillars were installed in a crystal configuration to maximize the trapping efficiency if the offset is $b_1/2$. Aligned configuration was able to trap only few beads. During the sensor experiment, we found that the pillar resulted in significantly difference electrokinetic responses.

On the main microchannel, we added side microchannels on both sides of the main- and buffer-microchannel for the easiness of the experiment[58]. Meniscus at air-liquid interface prevents the fluid leak through the valve. The meniscus can resist the amplified electrokinetic flow rate by ICP, while external high pressure can destruct the meniscus. Thus, one can repeat ICP experiment in a short time (<15 minutes) due to easy flushing the microchannel and obtain an identical electrokinetic response as similar as micro-nano-microchannel device (i.e. dead-end device). The dimensions of the microchannel were as following; the main- and buffer-microchannel: 200 μm width × 15 μm depth × 6 mm length; the air valve microchannel: 15 μm width × 15 μm depth. General PDMS fabrication steps were used to fabricate the device[59]. The Nafion nanoporous membrane was patterned on the glass substrate based on surface patterning method[60]. Simply, Nafion was patterned using a straight microchannel (200 μm width × 50 μm depth) on a glass side, and the PDMS piece of main microchannel was irreversibly bonded in the middle by plasma bonder (CuteMP, Femto Science, Korea) to a designated position on top of the Nafion-patterned glass.

**2.2.2. Chemical preparation**

For concentration visualization of the experiments, we only used fluorescent dye without any background electrolyte. Pure solid type of Alexa Flour 488 (Sigma Aldrich, m.w. 643.41 g/mol) of 1 mg was mixed with DI water of 1.55 mL to obtain 1mM solution. Due to the absence of background electrolyte, fluorescent molecules virtually acted like a major carrier in this experiment. For flow field tracking experiment, KCl solution (Sigma-Aldrich, USA) at 1 mM with ultra-sonicated canola-oil droplets. We have employed sonicated canola oil droplet whose zeta potential is below -10 mV. Thus, electroosmotic mobility / electrophoretic mobility < 10 because zeta potential of PDMS wall is approximately -100 mV. The low zeta potential allowed the droplet to enter the ion depletion zone. Highly charged tracers such as fluorescent particles can not enter the ion depletion zone. There is an electro-neutral particle, but its size is too big to be inserted into a microchannel. To prepare the oil droplet, a bottle of canola oil was purchased (Lotte Food, Korea) and the oil was ultra-sonicated for 15 minutes at 40 kHz condition. Then the oil was broken into the micro droplets whose diameter is less than 2 μm. The oil droplets are bigger than conventional small tracers, but we must use the oil droplet because of the low surface charge. For I-V measurement, 1 mM of KCl solution without dye and oil droplet was used.

### 2.2.3. Experimental setup

In the microfluidic experiment, the concentration profile inside the ICP layer and the tracers near the pillar array were imaged by an inverted fluorescent microscope (IX53, Olympus) and the CellSens program. Using Ag/AgCl electrodes, external electric field was applied by a source measure unit (Keithley 236, USA). Current-voltage response were cyclically measured in the range from 0 V to 10 V (0V → 10V → 0V) at ±0.2 V/30 sec. The current values were automatically recorded at every step by customized LabView program.

## 3. Results and Discussions

### 3.1 SC and EOF in crystal dielectric pillar configuration

In order to investigate the effect of dielectric pillars inside IDZ, one should determine the number of columns of the pillar in the device. An electric potential is applied to the main microchannel and the buffer microchannel is grounded so that IDZ is formed at the main microchannel where the pillars are installed. Detailed experimental methods were described in Methods section. Overlimiting conductance increases and saturates as the number of dielectric pillar array increases (See supporting information). This is attributed to the AIEZ and electroosmotic flow can enhance overlimiting conductance, but the dielectric pillars outside IDZ would merely affect to overlimiting conductance. Thus, we set the number of dielectric pillar array (9 rows and 12 columns) to let the IDZ remain inside the dielectric pillar arrays.

The first test configuration is a crystal configuration; *i.e.* pillars with a radius of 15 μm are alternatively positioned as shown in Figure 6. The simulation and experimental result of concentration profile inside IDZ are shown in Figure 6(a). Note that only SC and EOF are considered in the simulation. Detailed numerical methods were described in Methods section. In the numerical contour plot (Figure 6(a)), AIEZ can be clearly seen in the front of each pillar, especially the first column from the membrane. Thus, a concentration gradient inside IDZ in this configuration is largely deformed due to the formation of AIEZs. While the deformed concentration gradient was observed in the experiment as well, AIEZ or AIDZ are weakly observed (Figure 6(a): Experiment). Because the formation of AIEZs is occurred due to EOF as well, the simulated flow field and experiment tracking are shown in Figure 6(b). As expected, strong and localized vortices are formed counterclockwise at the top and clockwise at the bottom of each pillar to satisfy the continuity condition. Since EOF is surface driven flow, the speed of vortex is non-uniform (faster in the vicinity of a pillar and slower between pillars).

This is clearly observed in the flow field tracking experiment (Figure 6(b): Experiment). Micro oil droplets inside IDZ initially are pushed back toward the reservoir along with the expansion of IDZ. Momentarily, the droplets start to be trapped in a vortex at high speed near the pillar and low speed in the bulk. Note that not every dielectric pillar is able to capture the tracer because most of the tracers escape from IDZ even oil has a low surface charge. As shown in this flow field tracking, each vortex rotates locally so that EOF contributes to the ion transport only within the individual pillar. See supporting video for the experimental concentration profile (video 1) and flow field tracking (video 2) in crystal configuration. Thus, one can expect SC and EOF almost equally contribute the ion current (or current density) depending on electrokinetic parameters. To divide each contribution, a numerical simulation is conducted as shown in Figure 6(c). Each plot represents the EOF only- (red dash dot line) and SC only- (blue dash line) contribution and overall current density (black solid line) as a function of an applied voltage. Whole governing equations and boundary conditions were solved for the case of SC+EOF. In the case of EOF only, $Du$ was set to be 0, because $Du$ is the ratio of surface conduction and bulk conduction. For the case of SC only, $Pe^0 = 0$ led to convection-free ionic transportation so that there was only a surface conduction effect. Note that "SC+EOF" is not a simple linear sum of SC and EOF due to nonlinearity. The tile of I-V plots for various surface charge density and Debye length are presented in supporting information. The case of $\lambda_0 / a = 0.01$ represents the situations such as 1 μm radius pillar with 1 mM electrolyte (10 nm Debye length) or 100 nm radius pillar with 100 mM electrolyte (1 nm Debye length). Based on our calculation, the dominant mechanism to contribute to overlimiting current regime is surface conduction in the case of this pillar system. Especially, the former situation often corresponds to a biosensor platform with diluted target samples[57, 61]. Better stability and higher preconcentration efficiency would be achieved using the pillar structures based on our presenting results. Nonetheless, SC and EOF are almost equally contributed as compared to the aligned configuration which will be demonstrated in the next section.

**3.2 SC and EOF in aligned dielectric pillar configuration**

The second test type is an aligned configuration. As demonstrated in the previous section, the computed and visualized concentration profiles are shown in Figure 7(a). In this configuration, the concentration gradient in IDZ is more deformed than the crystal configuration. While IDZ still exist between each row of pillars, AIEZs are connected along each row of pillars. Thus, more charge-carriers in a stationary bulk can penetrate into IDZ along with the aligned AIEZ around the dielectric pillars. The individual AIEZ can be observed not only in the computational map but also in the experimentally visualized map. This is attributed to the inter-pillar convection due to EOF. As shown in Figure 7(b), the vortices around each dielectric pillar are horizontally inter-connected to form an elongated loop. See supporting video for the experimental concentration profile (video 1) and flow field tracking (video 2) in an aligned configuration. Similar to the crystal configuration, the speed of tracer is accelerated only when it passed around the pillar. However, images were in and out of focus, when the oil droplets were bouncing back and forth between columns in the aligned configuration, which indicated the droplet moved in $z$-direction as well. Furthermore, sinusoidal motion between the rows of pillars represented the elongated loop. From these observations, we presumably conclude that the contribution of EOF to the ion transportation would be larger in the aligned configuration than in the crystal configuration. The tile plots of electrokinetic parameters shown in Figure 7(c) support this observation (Also see supporting information for full tile plots). Noteworthily, the contribution of EOF (red dash dot line) is predominant in most cases of the aligned configuration, while one of SC (blue dot line) in this aligned configuration is almost identical to that of the crystal configuration.

**3.3 Overlimiting conductance enhancement by the dielectric pillars**

We are expecting significant enhancement of the ion transport by the microstructures if the scale of the device becomes submillimeter scale as mentioned. As shown in Figure 8, the ion transportation (*i.e.* overlimiting conductance) significantly increased even with the presence of the dielectric pillar arrays. AIEZs formed around the dielectric pillar array bring distant ion carriers closer to the membrane in this micro/nanofluidic device where low bulk concentration and the high electric field is imposed [62, 63]. This is directly associated with overlimiting conductance. The role of SC, EOF and EOI become distinguishable in overlimiting current regime as several literatures predicted and verified[5, 6, 20, 64-68]. This means I-V responses should be identical in under limiting current regime, especially when the surface area (for the same surface conduction) and cross-sectional area (for the same bulk conduction) of microchannel are kept to be constant. Thus, an invisible difference in the I-V curve in Figure 8 is correct result, because we design the same surface area and cross-sectional area of microchannels for each configuration.

# 4. Conclusions

Microstructure such as dielectric spacers used in conventional macroscale electrodialysis desalination systems have a negative impact on the perm-selective ion transport through nanoporous membrane due to the current shadow effect[13, 69]. The spacers lead to high electrical resistance and, therefore, to worse power consumption because the ions itself do not migrate where the spacers are located. Thus, in this study, we investigated the electrokinetic effects of the dielectric micro-pillar (which mimicked spacers in electrodialysis device) in a micro confinement environment by multi-physics simulation and micro/nanofluidic experiments. The microfluidic confinement would enhance the ion transportation since the length scale of such system is typically within the SC or EOF dominant environment[5, 20, 23], while macroscale electrodialysis system is in the EOI regime[30, 70, 71]. We numerically and experimentally confirmed that the combined contribution of SC and EOF caused by the dielectric pillars enhances the ion transport to the nanoporous membrane by forming multiple AIEZs, which is directly associated with OLC. Moreover, two alignment type of dielectric pillar array; staggered and aligned configuration, was rigorously investigated for how the physical structure can affect to the ion transportation. While vortical flow is isolated in staggered configuration, it can be elongated in the aligned configuration so that the latter configuration has higher ion transport efficiency, leading to higher OLC. Conclusively, the fundamental findings of the electrokinetic effects of the dielectric pillars in this study teach us that one should consider not only well-known electrokinetic parameters such as surface charge and bulk concentration but also the physical configuration of micro-confinement. These results would be an effective mean for designing an efficient electrochemical membrane platform including micro/nanofluidic devices.

# Acknowledgements

This work is supported by the Basic Research Laboratory Project (NRF-2018R1A4A1022513), Basic Science Research Program (2016R1A6A3A11930759) and the Center for Integrated Smart Sensor funded as Global Frontier Project (CISS- 2011-0031870) by the Ministry of Science and ICT and Korean Health Technology RND project from the Ministry of Health and Welfare Republic of Korea (HI13C1468). Also, all authors acknowledged the financial supports from BK21 Plus program of the Creative Research Engineer Development IT, Seoul National University.

# Additional Information

**Competing interests**

The authors declare no competing interests (both financial and non-financial).

# Author Contributions

K. Huh conducted the main experiment. S.-Y Yang fabricated the micro/nanofluidic device, J. S. Park conducted 3-dimensional droplet tracking and J. A. Lee supported numerical simulation. H. Lee analyzed the system numerically. H. Lee and S. J. Kim supervised the project. All authors wrote the manuscript.

# FIGURE CAPTIONS

**Figure 1.** (a) Preliminary experimental current-voltage characteristics of micro/nanofluidic device with and without micropillar. (b) Schematic diagram of two mechanisms of perm-selective ion transportation; (i) surface-driven electroosmosis and (ii) surface conduction through electrical double layer on the dielectric pillar. (c) Schematics representations of ion depletion zone (IDZ), ion enrichment zone (IEZ), additional ion depletion zone (AIDZ), additional ion enrichment zone (AIEZ) in micro/nanofluidic platform.

**Figure 2.** Schematic domain of electrochemical membrane system with dielectric pillar array.

**Figure 3.** Schematic diagram of computational domain using global (Cartesian) and local (cylindrical) coordinate system for (a) crystal and (b) aligned configuration.

**Figure 4.** Experimentally measured current-voltage relations for no pillar (rectangle), crystal (triangle) and aligned configuration (circle). Numerical simulation of concentration distribution at $Pe^0 = 0$ and $Pe^0 \neq 0$ of (a) crystal and (b) aligned configuration.

**Figure 5.** Microscopic images of fabricated micro/nanofluidic device, integrated with the dielectric pillars on anodic side.

**Figure 6.** (a) Numerically simulated and experimental visualized deformed concentration profile of the crystal dielectric pillar configuration. The numerically obtained concentration profile is normalized by bulk concentration $c_0$. (b) the flow field simulation and flow tracking experimental results of the crystal configuration, respectively. The flow field strength is normalized by $U_0 = Pe^0 D / L$ in which $L$ is the distance from stationary bulk to the cation-selective surface. (c) Current-voltage characteristics depending on Debye length ($\lambda_D$) of the crystal configuration.

**Figure 7.** (a) Numerically simulated and experimental visualized deformed concentration profile of the aligned dielectric pillar configuration. (b) the flow field simulation and flow tracking experimental results of aligned configuration, respectively. (c) Current-voltage characteristics depending on Debye length of the aligned configuration.

**Figure 8.** Experimentally measured current-voltage relations for no pillar (rectangle), crystal (triangle) and aligned configuration (circle).

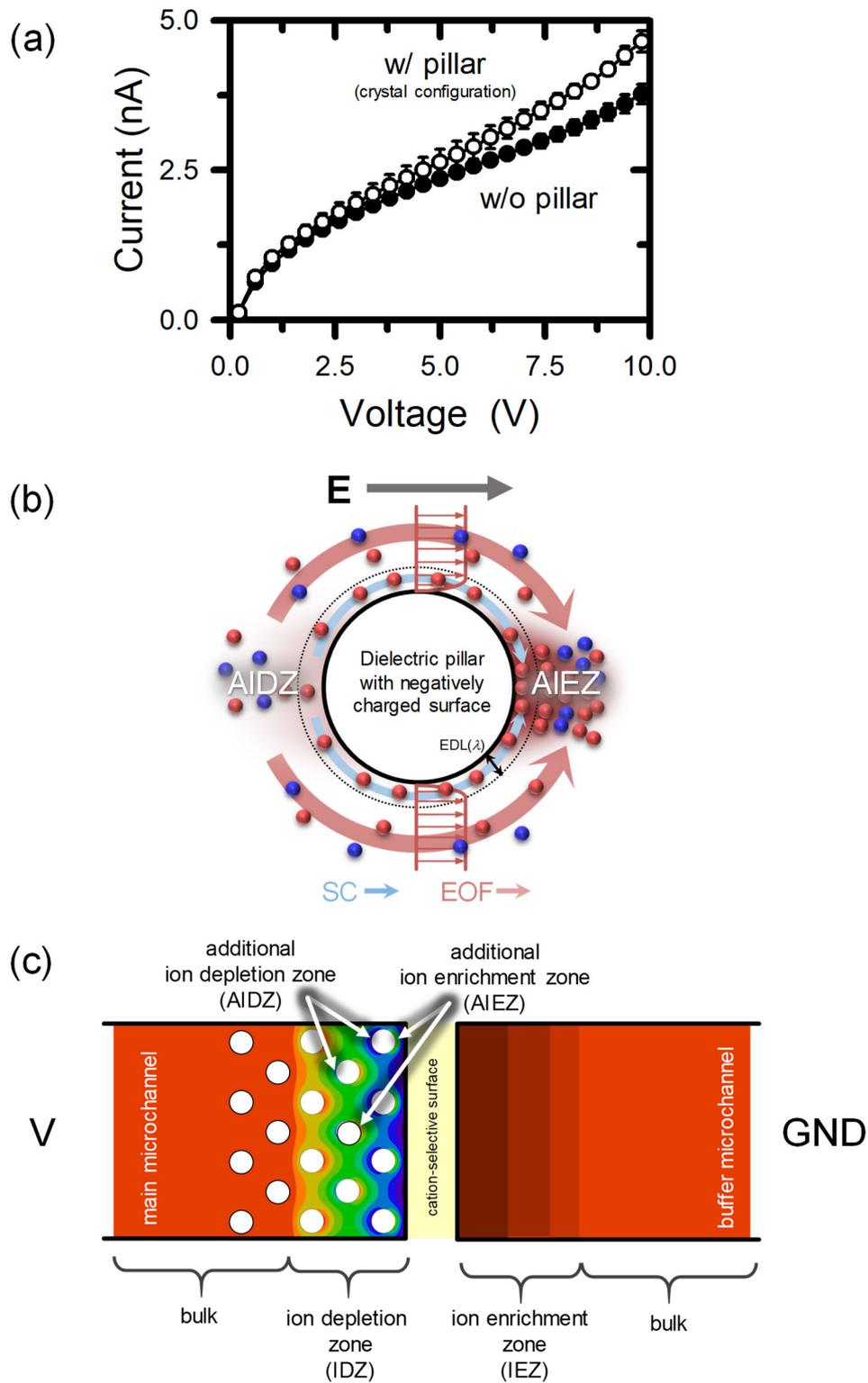

Figure 1

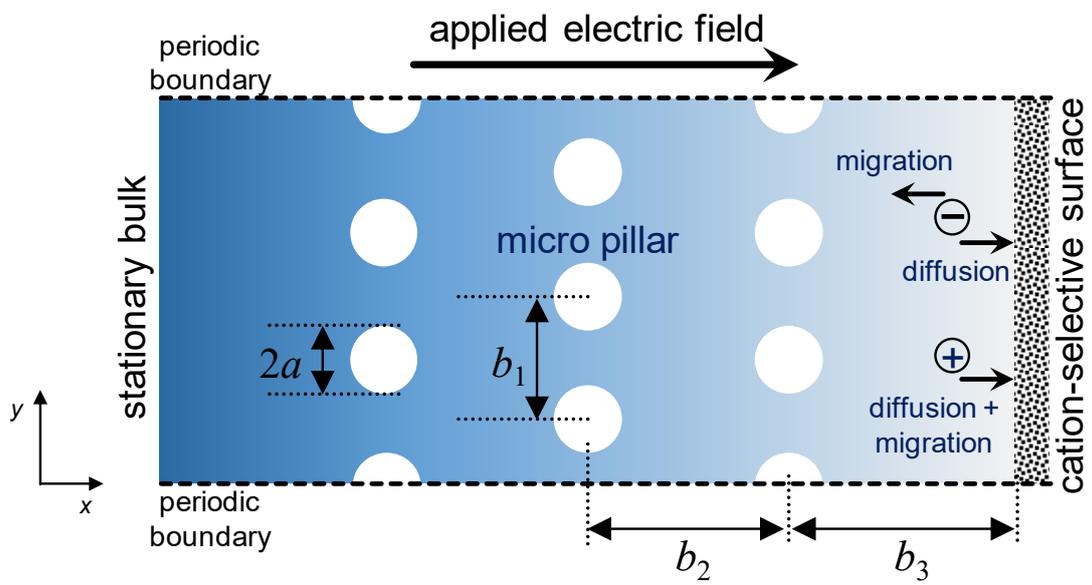

Figure 2

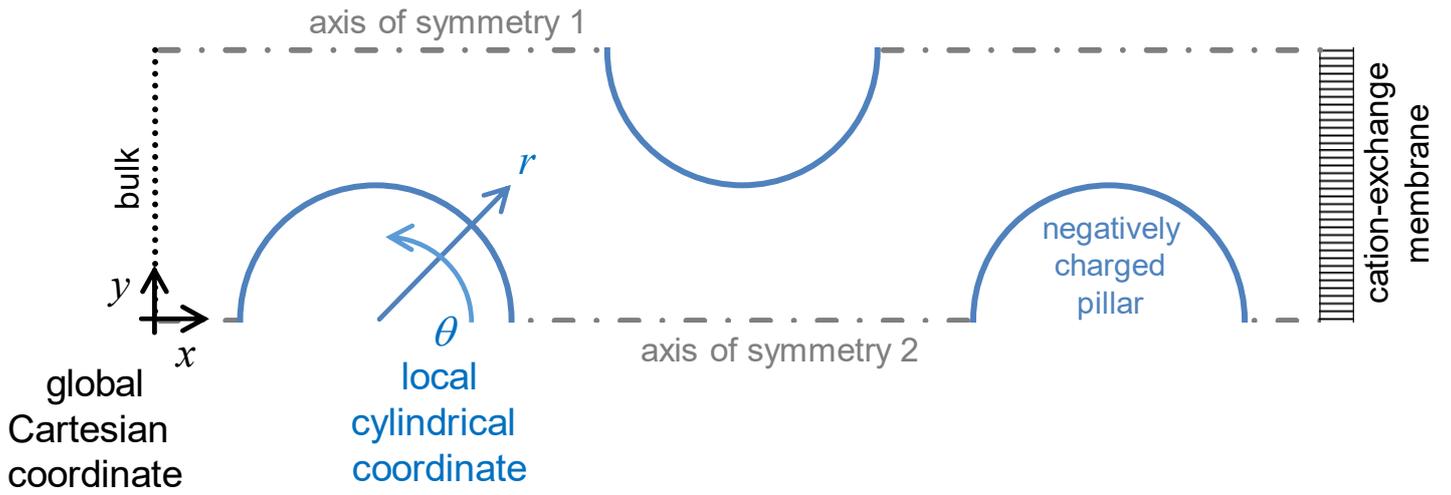

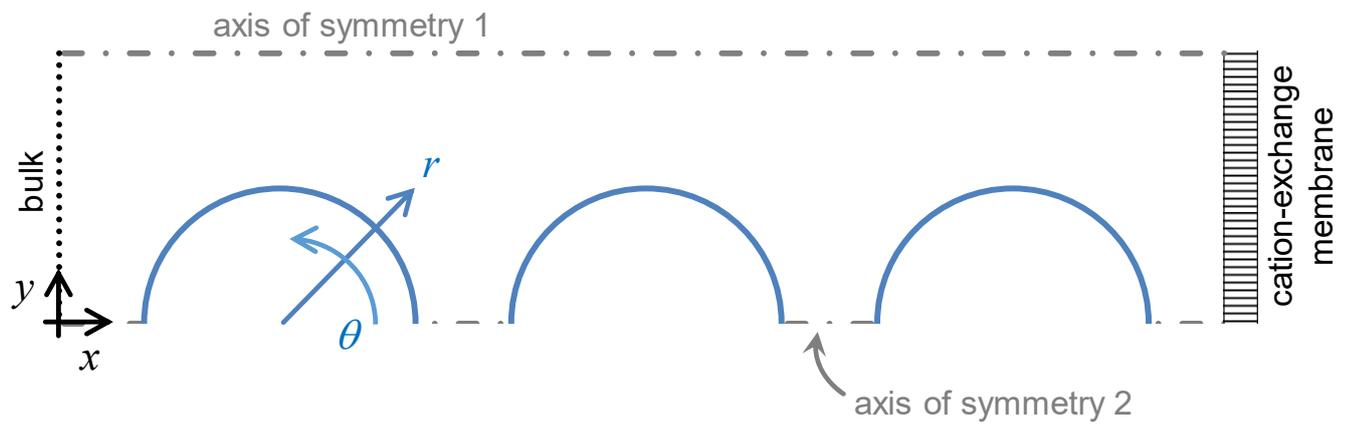

Figure 3

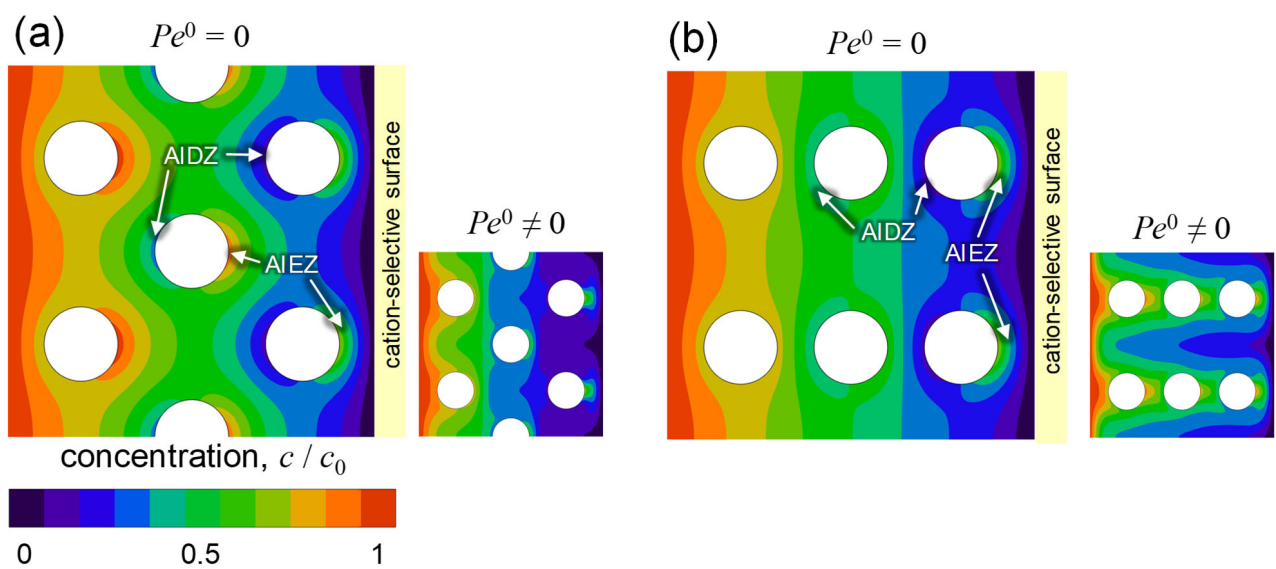

Figure 4

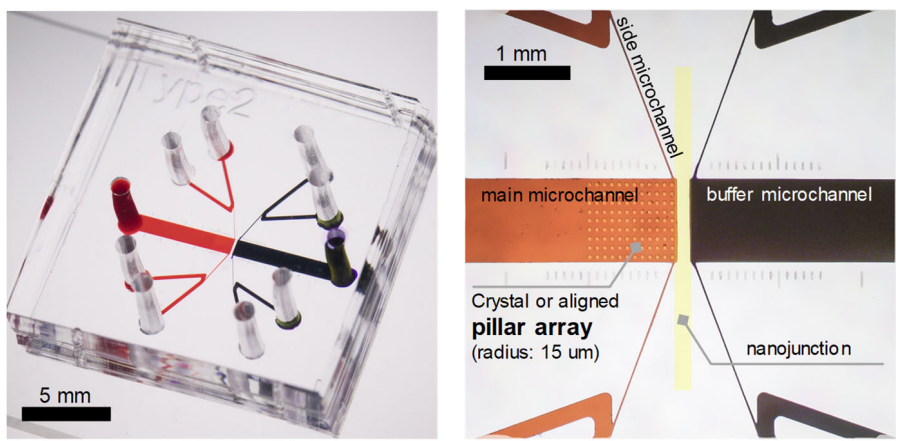

Figure 5

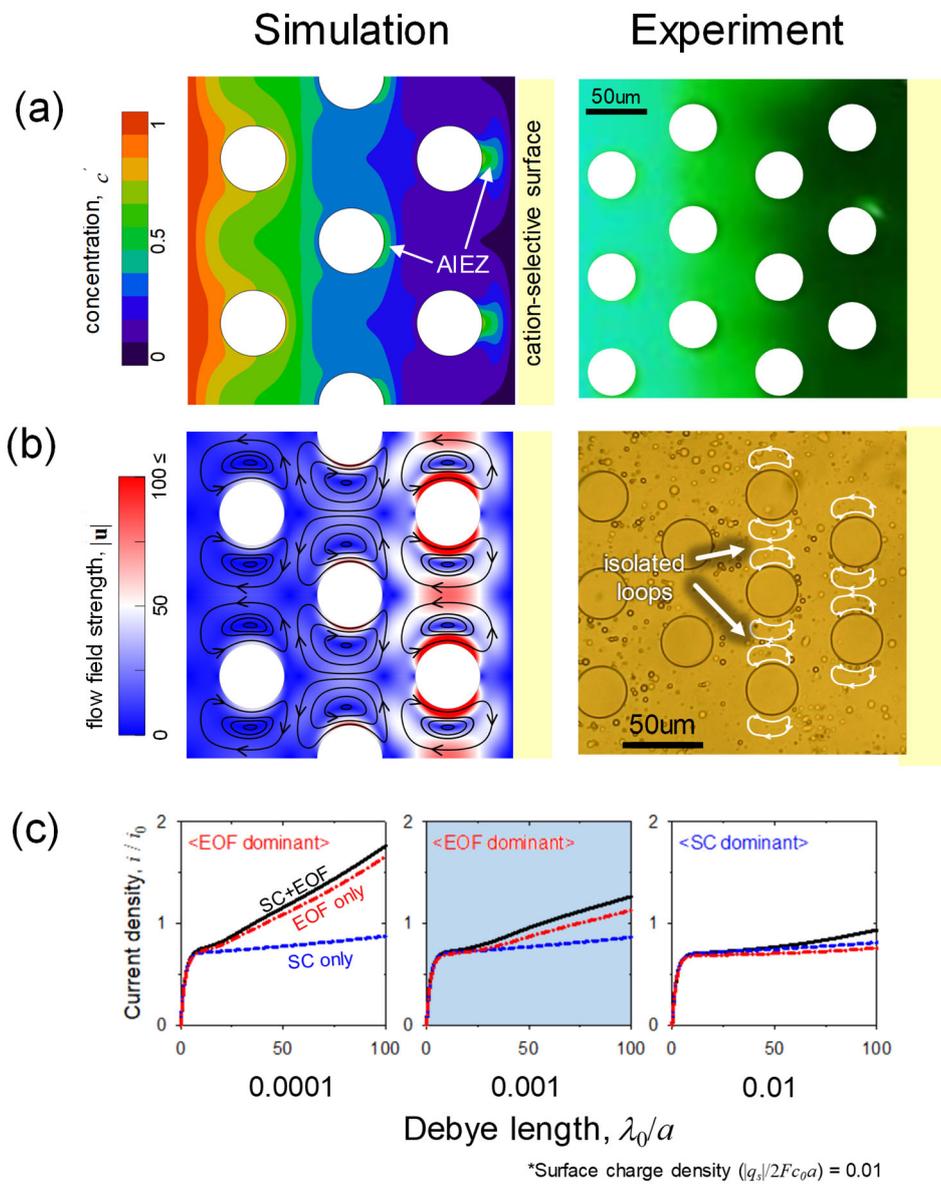

Figure 6

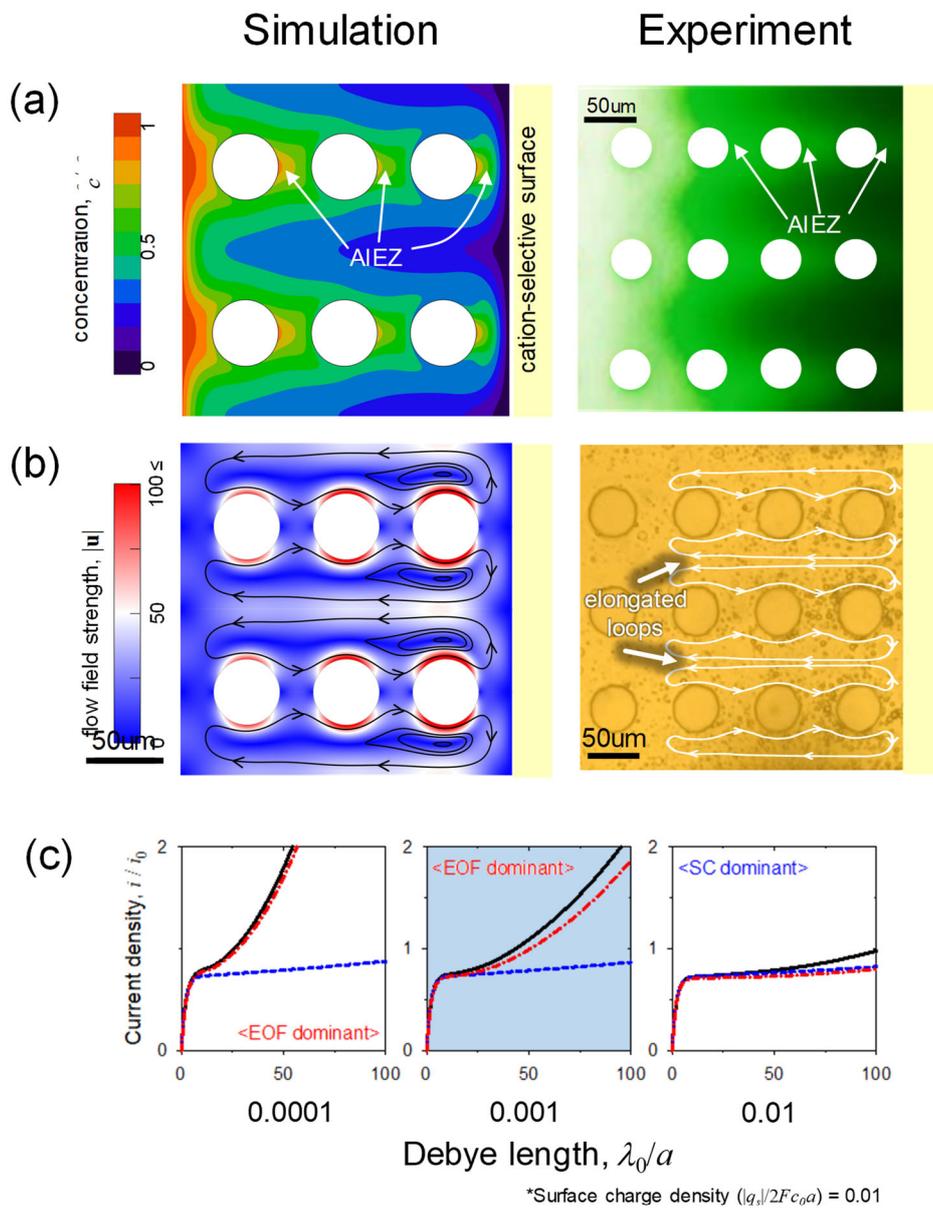

Figure 7

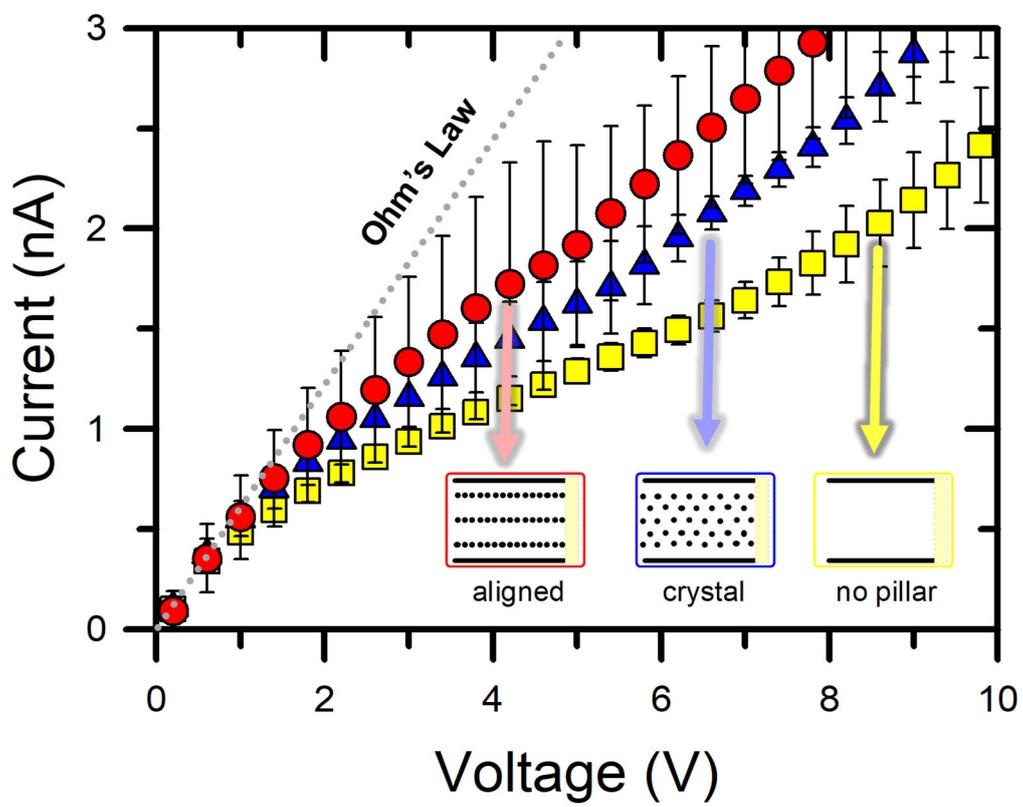

Figure 8